\documentclass[aps,twocolumn,pra,superscriptaddress,tightenlines]{revtex4}

\usepackage{epsfig,graphicx,times}
\usepackage{epstopdf}
\usepackage{amstext}
\usepackage{amsmath}
\usepackage{amssymb}
\usepackage{graphicx}
\usepackage{latexsym}
\usepackage{bm}
\usepackage[colorlinks,citecolor=blue, linkcolor=blue,hyperindex,CJKbookmarks,dvipdfm]{hyperref}

\begin{document}

\title{Generalized Su-Schrieffer-Heeger model in one dimensional
optomechanical arrays}
\author{Xun-Wei Xu}
\email{davidxu0816@163.com}
\affiliation{Department of Applied Physics, East China Jiaotong University, Nanchang,
330013, China}
\author{Yan-Jun Zhao}
\affiliation{Beijing National Laboratory for Condensed Matter Physics, Institute of
Physics, Chinese Academy of Sciences, Beijing 100190, China}
\affiliation{School of Physical Sciences, University of Chinese Academy of Sciences,
Beijing 100190, China}
\author{Hui Wang}
\affiliation{Department of Mechanical Engineering, National University of Singapore,
Singapore 117576}
\author{Ai-Xi Chen}
\email{aixichen@ecjtu.edu.cn}
\affiliation{Department of Physics, Zhejiang Sci-Tech University, Hangzhou 310018, China}
\affiliation{Department of Applied Physics, East China Jiaotong University, Nanchang,
330013, China}
\author{Yu-xi Liu}
\affiliation{Institute of Microelectronics, Tsinghua University, Beijing 100084, China}
\affiliation{Beijing National Research Center for Information Science and Technology (BNRist), Beijing 100084, China}
\date{\today }

\begin{abstract}
We propose an implementation of a generalized Su-Schrieffer-Heeger (SSH)
model based on optomechanical arrays. The topological properties of the
generalized SSH model depend on the effective optomechanical interactions
enhanced by strong driving optical fields. Three phases including one trivial
and two distinct topological phases are found in the generalized SSH model.
The phase transition can be observed by turning the strengths and phases of
the effective optomechanical interactions via adjusting the external driving fields. Moreover, four types of edge
states can be created in generalized SSH model of an open chain under
single-particle excitation, and the dynamical behaviors of the excitation in the open chain are related to the topological properties under the periodic boundary condition. We show that the edge states can be pumped adiabatically
along the optomechanical arrays by periodically modulating the amplitude and frequency of
the driving fields.
The generalized SSH model based on the optomechanical arrays
provides us a tunable platform to engineer topological phases for photons
and phonons, which may have potential applications in controlling the
transport of photons and phonons.
\end{abstract}

\maketitle


\section{Introduction}

In the past decades, rapid progress has been made in the field of
optomechanical systems, in which a cavity mode is coupled to a mechanical
mode via radiation pressure or optical gradient forces (for reviews, see
Refs.~\cite%
{KippenbergSci08,MarquardtPhy09,AspelmeyerPT12,AspelmeyerARX13,MetcalfeAPR14,YLLiuCPB18}%
). With the advance in technology and the requirements for providing new
functionality, the focus has been moved on from the simplest optomechanical
systems, based on a single mechanical mode coupled to a single optical mode,
to more complex setups which contain a few mechanical and optical modes, and
are designed as a periodic arrangement of optomechanical systems, i.e.,
optomechanical arrays. Based on the current condition of experiment and
technology, optomechanical arrays might be realized in coupled optical
microdisks~\cite{LinPRL09,MLiNatPhot09,WeisSci10,MZhangPRL15}, on-chip
superconducting circuit electromechanical cavity arrays~\cite%
{TeufelNat11,MasselNC12,PalomakiSci13,SuhSci14}, and optomechanical crystals~%
\cite{EichenfieldNat09,Safavi-NaeiniPRL14}.

In the past few years, quantum many-body effects in optomechanical arrays
have attracted considerable attentions. Optomechanical arrays with
parametric coupling between the mechanical mode and optical mode provide us
a controllable platform to simulate quantum many-body systems and manipulate
photons and phonons. Many interesting phenomena have been shown, such as
controllable photon propagation~\cite{ChangNJP11,WChenPRA14},
synchronization~\cite{HeinrichPRL11,LudwigPRL13,LauterPRE17}, artificial
magnetic fields for photons~\cite{SchmidtOptica15}, optically tunable
Dirac-type band structure~\cite{SchmidtNJP15}, Anderson localization of
hybrid photon-phonon excitations~\cite{RoqueNJP17,LLWanOE17}, and
Kuznetsov-Ma soliton~\cite{HXiongPRL17}. Besides these, another exciting
development is that the optomechanical arrays can be used to engineer
topological phases for both photons and phonons~\cite%
{PeanoPRX15,PeanoPRX16,PeanoNC16,MinkovOptica16,BrendelPRB18}.

Most of the optomechanical arrays, which are used to demonstrate different
topological phases and Chern insulators, are implemented in two-dimensional
optomechanical crystals. However, two dimensional systems are not a
necessary condition for engineering topological phases. The topological
properties of photons and phonons can also be implemented in a much simpler
platform, i.e., one-dimensional chain of optomechanical cavities. For
example, in a recent work, $Z_2$ topological insulators were simulated via a
one-dimensional optomechanical array~\cite{LQiOE17}.

It is generally known that the Su-Schrieffer-Heeger (SSH) model, introduced
from polyacetylene~\cite{HeegerRMP88}, is one of the simplest models to
demonstrate topological characters in one dimension, and now it has been
studied in many different settings, such as cold atoms and ions~\cite%
{BermudezNJP12,AtalaNatPhy13,GoldmanRPP14,JotzuNat14,DucaSci15}, optical
systems~\cite%
{LLuNatPhot2014,LLuNatPhys2014,KhanikaevNatPhot2014,XCSunPQE2017,OzawaArx18}%
, mechanical systems~\cite{FleuryAT15,HuberNatPhys16}, plasmonic systems~\cite{PoddubnyACS14,CWLingOE15,QChengLPR15,LGeOE15,CLiuOE18,DowningPRB17,Downingarx18}, and superconducting
circuit lattices~\cite%
{KochPRA10,NunnenkampNJP11,FMeiPRA15,FMeiQST16,ZHYangPRA16,TangpanitanonPRL16,XGuArx17}.
In addition, ladder systems, which consist of two or more coupled SSH chains, have been discussed to demonstrate richer topological quantum phases~\cite{ClayPRL05,XPLiNC13,ShimizuPRL15,CheonScience15,TZhangSR15,CLiPRB17}.

In this paper, we study the topological properties of a one-dimensional
optomechanical array, which can be mapped to the SSH model including three
complex hopping amplitudes~\cite{AsbothBook16}, called the generalized SSH
model. Differently from the standard SSH model, we find that there are three
phases in this generalized SSH model, one is trivial and other two are distinct.
It is worth mentioning that a SSH model consisting of three real hopping amplitudes was discussed in a recent reference~\cite{CLiPRB17}. Besides implementing with different systems~\cite{CLiPRB17}, we here show that the topological properties of the generalized SSH model depend on both the strengths and the phases of the hopping amplitudes, and topological phase transitions can be observed by tuning the strengths and phases of the effective optomechanical interactions via adjusting the external driving fields. What's more, four types of edge states can be found in generalized
SSH model of an open chain under single-particle excitation, and the dynamical behaviors of the excitation in the open chain are related to the topological properties under the periodic boundary condition. We show that the edge
states can be pumped adiabatically along the optomechanical arrays by
periodically modulating the amplitudes and frequencies of the driving fields.

The remainder of this paper is organized as follows. In Sec.~II, we show the
theoretical model of a generalized SSH model based on optomechanical arrays.
In Sec.~III, we study the topological properties of the generalized SSH
model and show that there are three phases, one is trivial and two are
distinct. Moreover, we show that phase transitions can be observed by tuning
the strengths of the optomechanical interactions. In Sec.~IV, four types of
edge states are introduced and the relation between the dynamical behaviors of single-particle excitation in the open chain and the topological properties under the periodic boundary condition are discussed. In Sec.~V, we
demonstrate that the edge states can be pumped adiabatically along the
optomechanical arrays by modulating the amplitudes and frequencies of the
driving fields periodically. Finally, the results are summarized in Sec.~VI.

\section{Theoretical model}

\begin{figure}[tbp]
\includegraphics[bb=30 440 588 693, width=8.5 cm, clip]{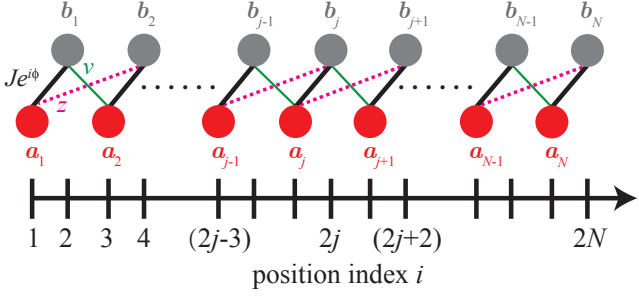}
\caption{(Color online) Schematic diagram of a generalized SSH model based
on an optomechanical array. $a_{j}$ and $b_{j}$ denote the cavity and
mechanical modes respectively, and they are coupled with three different
hopping amplitudes $Je^{i\protect\phi}$ (bold black lines), $v$ (thin green
lines) and $z$ (red dashed lines). The optomechanical array can be
implemented by an array of coupled optical microdisks~\protect\cite%
{LinPRL09,MLiNatPhot09,WeisSci10,MZhangPRL15}, on-chip superconducting
circuit electromechanical cavities~\protect\cite%
{TeufelNat11,MasselNC12,PalomakiSci13,SuhSci14}, and optomechanical crystals~%
\protect\cite{EichenfieldNat09,Safavi-NaeiniPRL14}.}
\label{fig1}
\end{figure}

We propose to implement a generalized SSH model by an optomechanical array
with $N$ cavity modes and $N$ mechanical modes, which are coupled only by
optomechanical interactions, without hopping of photons (or phonons) between
neighboring cavity modes (mechanical modes). The Hamiltonian of the
optomechanical array is ($\hbar =1$)
\begin{equation}
H_{0}=\sum_{j=1}^{N}\left( \omega _{c,j}a_{j}^{\dag }a_{j}+\omega
_{m,j}b_{j}^{\dag }b_{j}+H_{j}\right) ,  \label{Eq1}
\end{equation}%
with
\begin{eqnarray}
H_{1} &=&\left( g_{1,0}b_{1}^{\dag }+g_{1,+}b_{2}^{\dag }\right) a_{1}^{\dag
}a_{1}  \notag \\
&&+\left( \Omega _{1,0}e^{i\omega _{1,0}t}+\Omega _{1,+}e^{i\omega
_{1,+}t}\right) a_{1}+\mathrm{H.c.},  \label{Eq2}
\end{eqnarray}%
for the first cavity mode,
\begin{eqnarray}
H_{N} &=&\left( g_{N,0}b_{N}^{\dag }+g_{N,-}b_{N-1}^{\dag }\right)
a_{N}^{\dag }a_{N}  \notag \\
&&+\left( \Omega _{N,0}e^{i\omega _{N,0}t}+\Omega _{N,-}e^{i\omega
_{N,-}t}\right) a_{N}+\mathrm{H.c.},  \label{Eq3}
\end{eqnarray}%
for the last cavity mode, and
\begin{eqnarray}
H_{j} &=&\left( g_{j,0}b_{j}^{\dag }+g_{j,+}b_{j+1}^{\dag
}+g_{j,-}b_{j-1}^{\dag }\right) a_{j}^{\dag }a_{j}  \notag \\
&&+\left( \Omega _{j,0}e^{i\omega _{j,0}t}+\Omega _{j,+}e^{i\omega
_{j,+}t}+\Omega _{j,-}e^{i\omega _{j,-}t}\right) a_{j}  \notag \\
&&+\mathrm{H.c.},  \label{Eq4}
\end{eqnarray}%
for the $j$th cavity mode ($1<j<N$), where $a_{j}$ ($a_{j}^{\dag }$) is the
bosonic annihilation (creation) operator of the $j$th cavity mode ($%
j=1,2,\cdots ,N$) with resonant frequency $\omega _{c,j}$, $b_{j}$ ($%
b_{j}^{\dag }$) is the bosonic annihilation (creation) operator of the $j$th
mechanical mode with resonant frequency $\omega _{m,j}$, and $g_{j,0}$ (%
$g_{j,\pm }$) is the optomechanical coupling strength between the $j$th cavity
mode and the $j$th mechanical mode (the $(j\pm 1)$th mechanical mode). The
$j$th cavity mode is driven by a three-tone laser at frequencies $\omega _{j,0}=\omega
_{c,j}-\omega _{m,j}$ and $\omega _{j,\pm}=\omega _{c,j}-\omega _{m,j\pm 1}$ with amplitudes
$\Omega _{j,0}$ and $\Omega _{j,\pm}$ in the well resolved sidebands
regime ($\omega _{m,j}\gg \{\kappa _{j},\gamma _{j}\}$, where $\kappa _{j}$
is the decay rate of the $j$th cavity mode, and $\gamma _{j} $ is the damping
rate of the $j$th mechanical mode).

To linearize the Hamiltonians in Eqs.~(\ref{Eq2})-(\ref{Eq4}), we can write
the operator for each cavity modes as the sum of its classical mean value and quantum fluctuation
operator as $a_{j}\rightarrow \alpha _{j}(t)+a_{j}$.
The classical part $\alpha _{j}(t)$ can be given approximately as $\alpha
_{j}(t)\approx \alpha _{j,0}e^{i\omega _{j,0}t}+\alpha _{j,+}e^{i\omega
_{j,+}t}+\alpha _{j,-}e^{i\omega _{j,-}t}$, where the classical amplitude $%
\alpha _{j,0}$ ($\alpha _{j,\pm }$) is determined by solving the classical
equation of motion with only cavity drive $\Omega _{j,0}$ ($\Omega _{j,\pm
} $) at frequency $\omega _{j,0}$ ($\omega _{j,\pm}$).
We assume that: (i) \textrm{min}$\left[ \left\vert \alpha
_{j,0}\right\vert ,|\alpha _{j,\pm }|\right] \gg 1$, so that we can only
keep the first-order terms in the small quantum fluctuation operators; (ii)
\textrm{min}$\left[ \omega _{m,j},\left\vert \omega _{m,j}-\omega
_{m,j^{\prime }}\right\vert _{j^{\prime }=j\pm 1}\right] \gg \mathrm{max}%
\left[ \left\vert g_{j,0}\alpha _{j,0}\right\vert ,\left\vert g_{j,\pm
}\alpha _{j,\pm }\right\vert \right] $, such that the counter-rotating terms
can be neglected safely; in the interaction picture with respect to $H_{%
\mathrm{rot}}=\sum_{j=1}^{N}\left[ \omega _{c,j}a_{j}^{\dag }a_{j}+\omega
_{m,j}b_{j}^{\dag }b_{j}\right] $, the linearized form of the interaction
Hamiltonians in Eqs.~(\ref{Eq2})-(\ref{Eq4}) are obtained as%
\begin{equation}  \label{Eq5}
H_{1}=Je^{i\phi }a_{1}^{\dag }b_{1}+za_{1}^{\dag }b_{2}+\mathrm{H.c.},
\end{equation}%
\begin{equation}  \label{Eq6}
H_{N}=Je^{i\phi }a_{N}^{\dag }b_{N}+za_{N}^{\dag }b_{N-1}+\mathrm{H.c.},
\end{equation}%
\begin{equation}  \label{Eq7}
H_{j}=Je^{i\phi }a_{j}^{\dag }b_{j}+za_{j}^{\dag }b_{j+1}+va_{j}^{\dag
}b_{j-1}+\mathrm{H.c.},
\end{equation}%
where $Je^{i\phi }\equiv g_{j,0}\alpha _{j,0}$, $v\equiv g_{j,-}\alpha
_{j,-} $, and $z\equiv g_{j,+}\alpha _{j,+}$. Without loss of generality, we
assume that $J$, $v$ and $z$ are real coupling strengths and the effect of
the phase factor $\phi $ can be observed experimentally.

In summary, by substituting Eqs.~(\ref{Eq5})-(\ref{Eq7}) into Eq.~(\ref{Eq1}%
), the linearized Hamiltonian for the optomechanical array in the
interaction picture with respect to $H_{\mathrm{rot}}=\sum_{j=1}^{N}\left[
\omega _{c,j}a_{j}^{\dag }a_{j}+\omega _{m,j}b_{j}^{\dag }b_{j}\right] $ is
given by
\begin{equation}
H=\sum_{j=1}^{N}Je^{i\phi }a_{j}^{\dag }b_{j}+\sum_{j=1}^{N-1}\left(
va_{j+1}^{\dag }b_{j}+zb_{j+1}^{\dag }a_{j}\right) +\mathrm{H.c.},
\label{Eq8}
\end{equation}%
as schematically shown in Fig.~\ref{fig1}. The linearized Hamiltonian for
the optomechanical array shows a generalized SSH model with hopping
amplitude $z\neq 0$ between the $j $th cavity mode and $(j+1)$th mechanical mode. When
the coupling strength $z=0$, the Hamiltonian for the optomechanical array becomes the well-known
SSH model~\cite{AsbothBook16}.

\section{Topological phase transition}

\begin{figure}[tbp]
\includegraphics[bb=118 306 533 619, width=5 cm, clip]{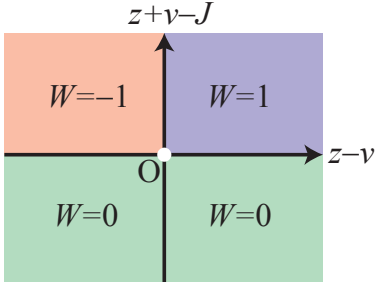}
\caption{(Color online) Phase diagram of the generalized SSH model with $\phi=0$. The
winding number is: $W=1$ if $z>v$ and $z+v>J$; $W=-1$ if $z<v$ and $%
z+v>J $; or $W=0$ if $z+v<J$.}
\label{fig2}
\end{figure}

To study the topological phase transition in the generalized SSH model, we
set a periodic boundary condition, so the linearized Hamiltonian for
one-dimensional optomechanical array can be redefined as
\begin{eqnarray}
H^{\prime }&=&\sum_{j=1}^{N}\left( Je^{i\phi }a_{j}^{\dag }b_{j}+va_{\left( j%
\text{ }\mathrm{mod}\text{ }N\right) +1}^{\dag }b_{j}+zb_{\left( j\text{ }%
\mathrm{mod}\text{ }N\right) +1}^{\dag }a_{j}\right) \nonumber\\
&& +\mathrm{H.c.},  \label{Eq9}
\end{eqnarray}%
where $\mathrm{mod}$ stands for the modular calculation. By using the for
Fourier transform for $a_{j}$ and $b_{j}$ as
\begin{equation}
\left(
\begin{array}{c}
a_{j} \\
b_{j}%
\end{array}%
\right) =\sum_{k}e^{-ikj}\left(
\begin{array}{c}
a_{k} \\
b_{k}%
\end{array}%
\right) ,
\end{equation}%
then the Hamiltonian in Eq.~(\ref{Eq9}) can be rewritten as%
\begin{equation}
H^{\prime }=\sum_{k}\left(
\begin{array}{cc}
a_{k}^{\dag } & b_{k}^{\dag }%
\end{array}%
\right) H^{\prime }\left( k\right) \left(
\begin{array}{c}
a_{k} \\
b_{k}%
\end{array}%
\right)
\end{equation}
with%
\begin{equation}
H^{\prime }\left( k\right) =\left(
\begin{array}{cc}
0 & h\left( k\right)  \\
h^{\ast }\left( k\right)  & 0%
\end{array}%
\right)
\end{equation}%
and%
\begin{equation}\label{eq09}
h\left( k\right) =Je^{i\phi }+ve^{-ik}+ze^{ik},
\end{equation}%
where $k$ is the wavenumber in the first Brillouin zone. The dispersion
relation of the generalized SSH model with periodic boundary conditions is
given by
\begin{equation}
E=\pm \left\vert Je^{i\phi }+ve^{-ik}+ze^{ik}\right\vert .
\end{equation}

The winding number, which characterizes the topological invariant of an
insulating Hamiltonian, is defined by~\cite{AsbothBook16}
\begin{equation}
W=\frac{1}{2\pi i}\int_{-\pi }^{\pi }dk\frac{d\ln h\left( k\right) }{dk}.
\end{equation}%
For the generalized SSH model with $h\left( k\right)$ given by Eq.~(\ref{eq09}), the winding number is either $0$ or $\pm 1$, depending on the parameters. When the phase factor $\phi=0$, see Fig.~\ref{fig2}, the winding number
is: (i) $W=1$ if $z>v$ and $z+v>J$; (ii) $W=-1$ if $z<v$ and $%
z+v>J$; (iii) $W=0$ if $z+v<J$.

The Hamiltonian for the generalized SSH model in the momentum space can be
written in an alternative form as
\begin{eqnarray}
H^{\prime }\left( k\right) &=&\mathbf{d}\left( k\right) \cdot \widehat{%
\mathbf{\sigma }}  \notag \\
&=&d_{x}\left( k\right) \widehat{\sigma }_{x}+d_{y}\left( k\right) \widehat{%
\sigma }_{y}+d_{z}\left( k\right) \widehat{\sigma }_{z},
\end{eqnarray}%
where $\widehat{\sigma }_{x}$, $\widehat{\sigma }_{y}$, and $\widehat{\sigma }_{z}$ are the Pauli matrices, and
\begin{equation}
d_{x}\left( k\right) =J\cos \phi +v\cos k+z\cos k,
\end{equation}%
\begin{equation}
d_{y}\left( k\right) =J\sin \phi -v\sin k+z\sin k,
\end{equation}%
\begin{equation}
d_{z}\left( k\right) =0.
\end{equation}%
The winding number can also be obtained graphically by counting the number
of times the loop winds around the origin of the $d_{x},d_{y}$ plane.

We show the dispersion relation and the path that the endpoint of the vector
$\mathbf{d}\left( k\right) $ traces out in Fig.~\ref{fig3} for $v>z$. As the
wavenumber runs through the Brillouin zone, $k=0\rightarrow 2\pi $, the path
that the endpoint of the vector $\mathbf{d}\left( k\right) $ is a closed
ellipse of long axis $v+z$ and short axis $|v-z|$ on the $d_{x},d_{y}$
plane, centered at $\left( J,0\right)$, and the endpoint rotates around the
origin clockwise. It is clear that the winding number is $W=-1$ when $z+v>J$
for $z<v$, and the winding number is $W=0$ when $z+v<J$.

Two more figures about the dispersion relation and the path that the
endpoint of the vector $\mathbf{d}\left( k\right) $ traces are shown in
Figs.~\ref{figA1} and \ref{figA2}, given in the Appendix. In Fig.~\ref{figA1}%
, as $v=z$, the path of the endpoint of the vector $\mathbf{d}\left(
k\right) $ becomes a straight line on the $d_{x}$-axis. In Fig.~\ref{figA2},
as $v<z$, similarly to the case for $v>z$, the path of the endpoint of the
vector $\mathbf{d}\left( k\right) $ is also a closed ellipse of long axis $%
v+z$ and short axis $|v-z|$ on the $d_{x},d_{y}$ plane, centered at $\left(
J,0\right)$. However, the endpoint rotates around the origin
counterclockwise for $v<z$. So we can conclude that the winding number is (i) $%
W=1$ when $z+v>J$ and $v<z$; (ii) $W=-1$ when $z+v>J$ and $%
v>z$; (iii) $W=0$ when $z+v<J$. These consist with the
results shown in Fig.~\ref{fig2}.

The above results are obtained under the condition for $\phi=0$.
In Fig.~\ref{fig4}, we show that the topological phase transition can be
induced by tuning the phase $\phi$. For $\phi\neq 0$, the path that the
endpoint of the vector $\mathbf{d}\left(k\right) $ is centered at $\left(
J\cos \phi ,J\sin \phi \right) $. The topological phase transition appears
when $E=0$, as shown in Fig.~\ref{fig4}(b), which gives the critical phase $%
\phi_c$ as
\begin{equation}
\tan \phi _{c}=\frac{z-v}{z+v}\tan k,
\end{equation}%
where
\begin{equation}
\cos ^{2}k=\frac{J^{2}-\left( z-v\right) ^{2}}{\left( z+v\right) ^{2}-\left(
z-v\right) ^{2}}.
\end{equation}%
As $v>z$, we have $W=-1$ for $\phi=0$ in Fig.~\ref{fig4}(a); we have $W=0$ for $\phi=0.1$ in Fig.~\ref%
{fig4}(c); the winding number is not well-defined in the critical point for $%
\phi=\phi_c\approx 0.04$ as shown in Fig.~%
\ref{fig4}(b). The paths [see Fig.~\ref{fig4}(d)-(f)] that the
endpoint of the vector $\mathbf{d}\left( k\right) $ traces, corresponding to dispersion relation [see Fig.~\ref{fig4}(a)-(c)] for $v<z$, show the phase transition from $W=1$ to $W=0$ by tuning $\phi$.

\begin{widetext}
\begin{figure*}[tbp]
\includegraphics[bb=9 336 585 568, width=15 cm, clip]{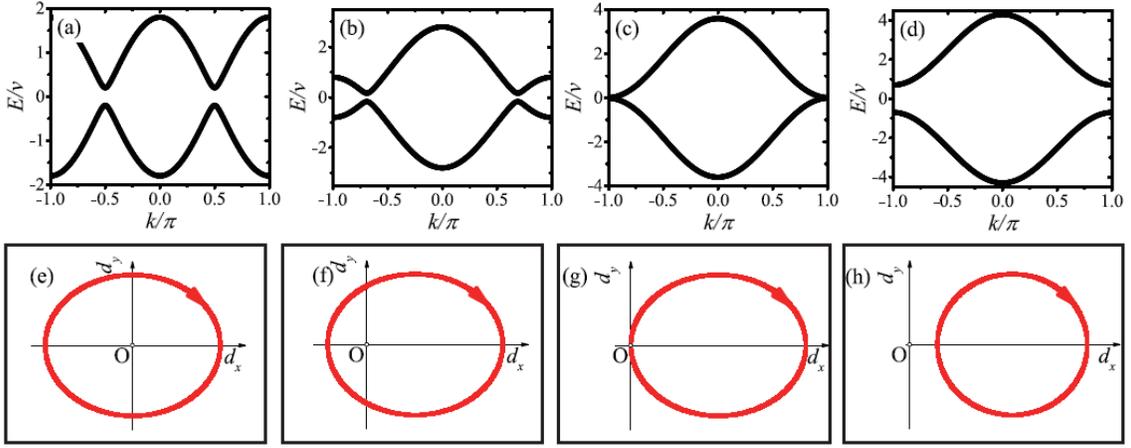}
\caption{(Color online) Dispersion relations of the generalized SSH model with different hopping amplitude $J$: (a) $J=0$; (b) $J/v=1$; (c) $J/v=1.8$; (d) $J/v=2.5$. (e)-(h) The paths of the endpoint of the vector $\textbf{d}(k)$ corresponding to (a)-(d) are shown on the $d_x,d_y$ plane as the wavenumber is sweeped across the Brillouin zone, $k=0\rightarrow 2\pi$. The other parameters are $z/v=0.8$ and $\phi=0$.}
\label{fig3}
\end{figure*}

\begin{figure*}[tbp]
\includegraphics[bb=14 272 578 569, width=11 cm, clip]{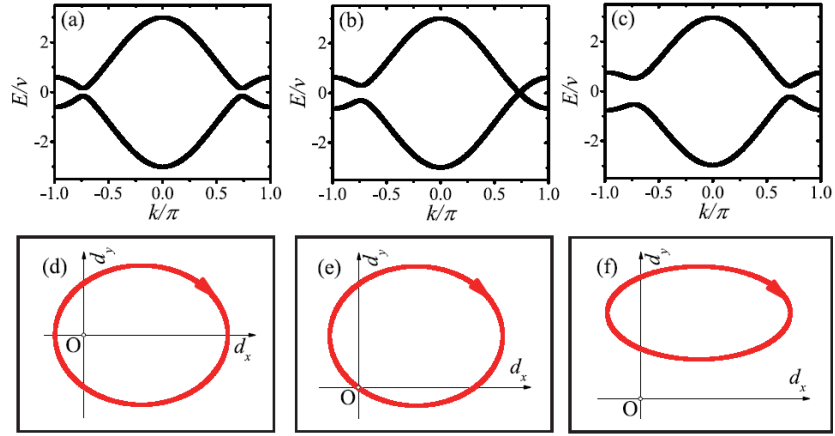}
\caption{(Color online) Dispersion relations of the generalized SSH model for hopping amplitude $J/v=1.2$ with different phase: (a) $\phi=0$; (b) $\phi=\phi_c\approx 0.04$; (c) $\phi=0.1$. (d)-(f) The paths of the endpoint of the vector $\textbf{d}(k)$, corresponding to (a)-(c), are shown on the $d_x,d_y$ plane as the wavenumber is sweeped across the Brillouin zone, $k=0\rightarrow 2\pi$. The other parameter is $z/v=0.8$.}
\label{fig4}
\end{figure*}

\end{widetext}

\section{Edge states}

We now study how to demonstrate topologically protected edge states in the
optomechanical array of an open chain under single-particle excitation. To
be specific, we choose an open chain of $N=8$ for the optomechanical array, then the wave function for the Hamiltanion in Eq.~(\ref{Eq8}) under single-particle excitation can be defined as
\begin{equation}
\left\vert \Psi \left( t\right) \right\rangle =\sum_{j=1}^{N}\left[
c_{2j-1}\left( t\right) \left\vert a_{j}\right\rangle +c_{2j}\left( t\right)
\left\vert b_{j}\right\rangle \right] ,
\end{equation}%
where $P_{2j-1}\left( t\right) =\left\vert c_{2j-1}\left( t\right)
\right\vert ^{2}$ and $P_{2j}\left( t\right) =\left\vert c_{2j}\left(
t\right) \right\vert ^{2}$ denote the occupying probabilities in the $j$th cavity mode
and $j$th mechanical mode, respectively. For simplicity, we define $%
P_{i}\left( t\right) =\left\vert c_{i}\left( t\right)\right\vert ^{2}$ with
the position index $i$ shown in Fig.~\ref{fig1}.

\begin{figure}[tbp]
\includegraphics[bb=44 214 546 623, width=8.5 cm, clip]{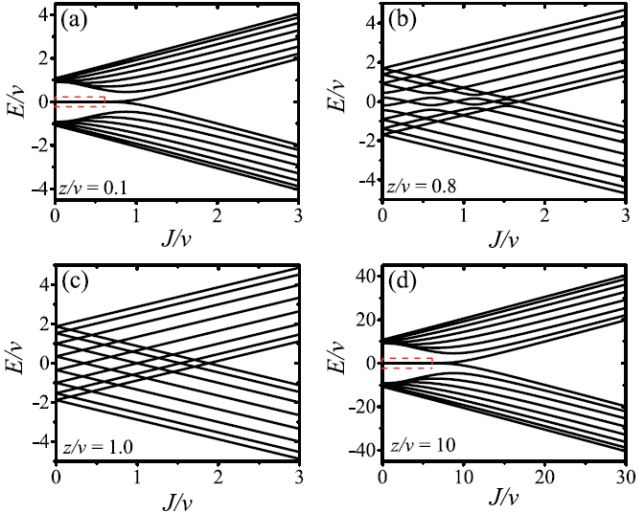}
\caption{(Color online) Energy spectrum of the open chain optomechanical array for $N = 8$ as
a function the intracell hopping amplitude $J$ for different intercell
hopping amplitude $z$: (a) $z/v=0.1$; (b) $z/v=0.8$; (c) $z/v=1.0$; (d) $z/v=10$.
The phase factor is $\protect\phi=0$.}
\label{fig5}
\end{figure}

The energy spectrum of a generalized SSH model with $N=8$ is shown in Fig.~%
\ref{fig5}. Under the conditions $z\ll v$ or $z\gg v$, see Figs.~\ref{fig5}%
(a) and \ref{fig5}(d), the energy spectrum of a generalized SSH model is
similar to that of the standard SSH model. However, due to the level crossings for
the nearest neighbor eigenmodes and avoided level crossings for the
next-nearest neighbor eigenmodes, there are four ($N/2$) degenerate points of the
zero-energy modes within the interval $0<J<z+v$, as shown in Figs.~\ref{fig6}%
(a) and \ref{fig6}(b), which are the local enlarged drawings of the boxes
with red dashed-line in Figs.~\ref{fig5}(a) and \ref{fig5}(d). When $z=0.8v$%
, as shown Fig.~\ref{fig5}(b), there are degenerate points for the nearest
neighbor eigenmodes and avoided level crossings for the next-nearest
neighbor eigenmodes. But when $z=v$, as shown Fig.~\ref{fig5}(c), all the
avoided level crossings for the next-nearest neighbor eigenmodes disappear.

The probability distributions of the eigenstates, corresponding to the
points marked out in Figs.~\ref{fig6}(a) and \ref{fig6}(b), are shown in
Figs.~\ref{fig6}(c)-\ref{fig6}(f). Apparently, the probability distributions of the eigenstates are localized. We define four edge states, i.e., left cavity (LC), left
mechniacal (LM), right cavity (RC), right mechniacal (RM) edge states, as%
\begin{equation}  \label{Eq21}
\left\vert LC\right\rangle =\sum_{j=1}^{N}c_{1}e^{-2\left( j-1\right) /\xi
}\left\vert a_{j}\right\rangle ,
\end{equation}
\begin{equation}  \label{Eq22}
\left\vert LM\right\rangle =\sum_{j=1}^{N}c_{2}e^{-2\left( j-1\right) /\xi
}\left\vert b_{j}\right\rangle ,
\end{equation}
\begin{equation}  \label{Eq23}
\left\vert RC\right\rangle =\sum_{j=1}^{N}c_{2N-1}e^{2\left( j-N\right) /\xi
}\left\vert a_{j}\right\rangle ,
\end{equation}
\begin{equation}  \label{Eq24}
\left\vert RM\right\rangle =\sum_{j=1}^{N}c_{2N}e^{2\left( j-N\right) /\xi
}\left\vert b_{j}\right\rangle,
\end{equation}
where $\xi>0$ is the localization length determined by the relative size of the hopping
amplitudes $v$, $z$, and $J$. When $v \gg z$ and $J \ll v+z$ (corresponding to the winding number $W=-1$ under the periodic boundary condition), as shown in Figs.~\ref{fig6}(c) and \ref{fig6}(d),
the edge states are the hybridized states of the LC edge state and RM edge state. A concise physical picture for the edge states with $v \gg z$ and $J \ll v+z$ is shown in Fig.~\ref{fig6}(g), where dimers are formed between $b_{j}$ and $a_{j+1}$, and $a_{1}$ and $b_{N}$ are isolated from the others. Similarly, when $v \ll z$ and $J \ll v+z$ (corresponding to the winding number $W=1$ under the periodic boundary condition), the edge states are the hybridized states of the LM edge state and RC edge state, as shown in Figs.~\ref{fig6}(e) and \ref{fig6}(f). The physical picture for the edge states with $v \ll z$ and $J \ll v+z$ is shown in Fig.~\ref{fig6}(h), where dimers are formed between $a_{j}$ and $b_{j+1}$, and $b_{1}$ and $a_{N}$ are isolated from the others.
When $J > v+z$ (corresponding to the winding number $W=0$ under the periodic boundary condition), dimers are formed between $a_{j}$ and $b_{j}$, and no modes are isolated from the others (the physical picture is not shown in the text). Thus, there is no edge states when $J > v+z$.

Figure~\ref{fig7} shows the time evolution of the probability distribution for the generalized SSH model with the open boundary condition (i.e., an open chain with $N=8$): (a)-(e) an excitation is injected at the first cavity mode
$P_{1}(0)=1$ for $z = v/10$, or (f)-(j) an excitation
is injected at the first mechanical mode $P_{2}(0)=1$ for $z=10 v$.
From these figures, we can see that:
(i) When $J=z \ll v$ or $J=v\ll z$, as shown in Figs.~\ref{fig7}(a) and \ref{fig7}(f), the excitation almost localizes in
the injected cavity or mechanical mode like a soliton for a long time.
With a larger value of $J$ (still with $J<z+v$), as shown in Figs.~\ref{fig7}(b) and \ref{fig7}(g), the excitation spreads to the nearest neighbor modes and the localization of the excitation becomes weaker.
(ii) When $J = z+v $ (i.e., the topological phase transition point), as shown in Figs.~\ref{fig7}(d) and \ref{fig7}(i), the excitation, oscillating like a soliton, travels in the open chain and reflects back at the ends.
The traveling speed (the period of oscillation) is dependent on the hopping amplitudes: the excitation
travels much faster in Fig.~\ref{fig7}(i) with $z/v=10$ and $J/v=11$ than in Fig.~%
\ref{fig7}(d) with $z/v=0.1$ and $J/v=1.1$.
As $J$ is away from the topological phase transition point ($J \neq z+v$), as shown in Figs.~\ref{fig7}(c) and \ref{fig7}(e) [or Figs.~\ref{fig7}(h) and \ref{fig7}(j)], the distribution of the traveling excitation disperses much faster than the case with $J = z+v $.
That is to say, the optimal condition for the excitation traveling like a soliton with less dispersion in the open chain is the system working at the critical point $J = z+v$.

\begin{widetext}
\begin{figure*}[tbp]
\includegraphics[bb=21 312 593 526, width=15 cm, clip]{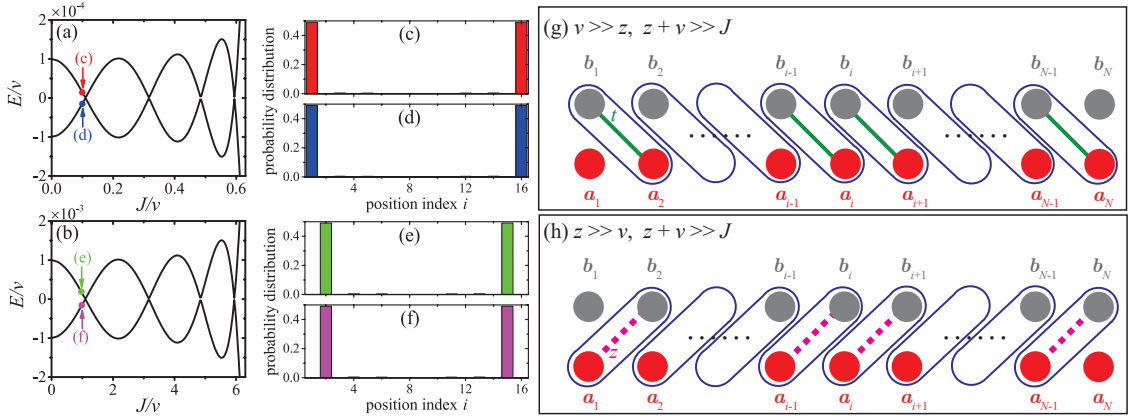}
\caption{(Color online) (a) and (b) are the local enlarged drawings of
\protect\ref{fig5}(a) and \protect\ref{fig5}(d). (c)-(f) show the
probability distributions of the eigenfunctions corresponding to the points
shown in (a) and (b). (g) Edge states appear at $a_{1}$ and $b_{N}$ corresponding to (c) and (d) for $v \gg z$ and $J \ll v+z$.
(h) Edge states appear at $b_{1}$ and $a_{N}$ corresponding to (e) and (f) for $z \gg v$ and $J \ll v+z$.}
\label{fig6}
\end{figure*}

\begin{figure*}[tbp]
\includegraphics[bb=48 324 555 546, width=15 cm, clip]{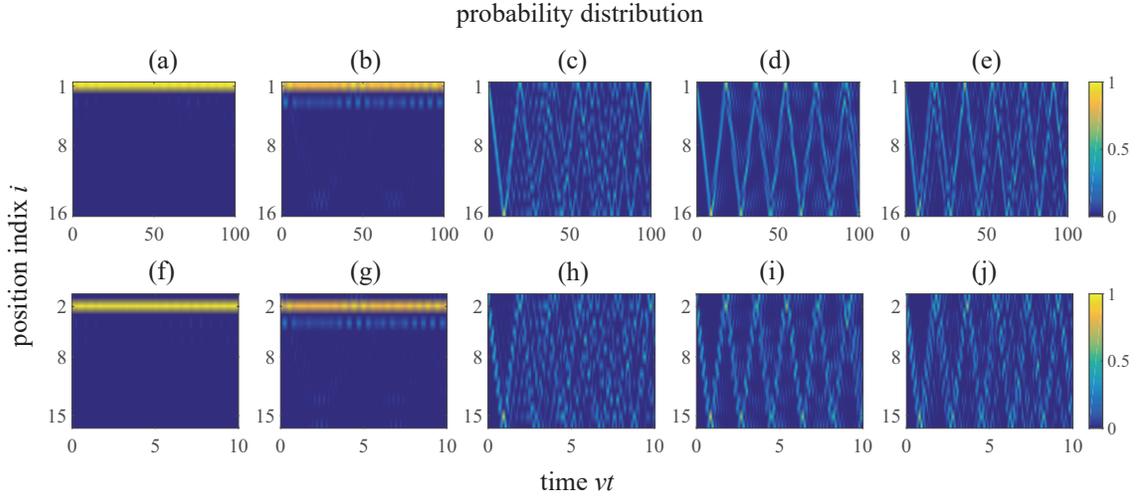}
\caption{(Color online) Time evolution of the probability distribution of
an open chain for $N=8$: (a)-(e) $z/v=0.1$ and $P_{1}(0)=1$, with (a) $J/v=0.1$, (b) $J/v=0.3$, (c) $J/v=1.0$, (d) $J/v=1.1$, (e) $J/v=1.2$; (f)-(j) $z/v=10$ and $P_{2}(0)=1$, with (f) $J/v=1$, (g) $J/v=3$, (h) $J/v=10$, (i) $J/v=11$, (j) $J/v=12$. The phase factor is $\protect\phi=0$.}
\label{fig7}
\end{figure*}

\end{widetext}

\section{Adiabatic particle pumping}

As shown in the previous section, once an excitation is injected at one
edge, it will stay there like a stationary state. However, we will show that it is
possible to transfer the edge states from one to another by adiabatic
pumping with periodically modulated optomechanical array with Hamiltonian as%
\begin{eqnarray}  \label{Eq25}
H &=&\sum_{j=1}^{N}\left[ u\left( t\right) a_{j}^{\dag }a_{j}+J\left(
t\right) a_{j}^{\dag }b_{j}\right]  \notag \\
&&+\sum_{j=1}^{N-1}\left[ v\left( t\right) a_{j+1}^{\dag }b_{j}+z\left(
t\right) b_{j+1}^{\dag }a_{j}\right] +\mathrm{H.c.},
\end{eqnarray}%
where $u\left( t\right) $ is the detuning between the cavity modes and
mechanical modes. The modulated optomechanical array can be realized by
modulating the frequencies and strengths of the driving fields.

\begin{figure}[tbp]
\includegraphics[bb=28 221 568 621, width=8.5 cm, clip]{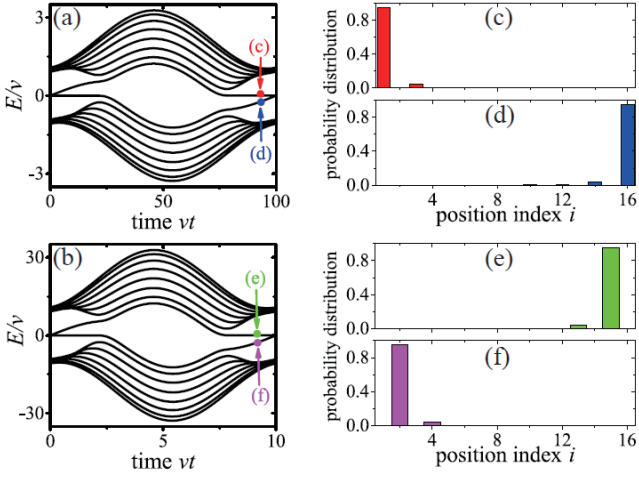}
\caption{(Color online) Instantaneous spectrum of the open chain optomechanical array for $N=8$. The time dependent pump sequence is defined in Eq.~(\protect
\ref{Eq26}), with (a) $z/v=0.1$, $A/v=1.1$, and $\omega/v=2\pi/100$; (b) $z/v=10$, $A/v=11$, and
$\omega/v=2\pi/10$. (c)-(f) show the probability distributions of the edge states
corresponding to the points shown in (a) and (b). The phase factor is $\protect\phi=0$.}
\label{fig8}
\end{figure}

\begin{figure}[tbp]
\includegraphics[bb=72 393 543 588, width=8.5 cm, clip]{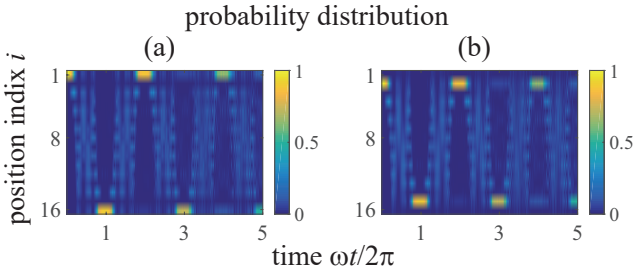}
\caption{(Color online) Time evolution of the probability distribution for
an open chain of $N=8$. The time dependent pump sequence is defined in Eq.~(%
\protect\ref{Eq26}), with (a) $z/v=0.1$, $A/v=1.1$, $\omega/v=2\pi/100$, and $P_{1}(0)=1$;
(b) $z/v=10$, $A/v=11$, $\omega/v=2\pi/10$, and $P_{2}(0)=1$. The phase factor is $\protect\phi=0$.}
\label{fig9}
\end{figure}

First, we consider a smooth modulation sequence as
\begin{equation}  \label{Eq26}
\left\{
\begin{array}{c}
u\left( t\right) =\frac{A}{2}\sin \left( \omega t\right) , \\
J\left( t\right) =A\left[ 1-\cos \left( \omega t\right) \right] , \\
v\left( t\right) =\mathrm{const.}, \\
z\left( t\right) =\mathrm{const.},%
\end{array}%
\right.
\end{equation}
where, $u(t)$ and $J(t)$ are modulated periodically with amplitudes $A/2$
and $A$, and frequency $\omega$; $u(t)$ and $J(t)$ are maintained constant.

The instantaneous spectrum of the Hamiltonian in Eq.~(\ref{Eq25}) with time
dependent pump sequence defined in Eq.~(\ref{Eq26}) is shown in Figs.~\ref%
{fig8}(a) and \ref{fig8}(b). In Figs.~\ref{fig8}(c)-\ref{fig8}(f), the
probability distributions of eigenstates corresponding to the points shown
in Figs.~\ref{fig8}(a) and \ref{fig8}(b), are approximately the edge states
defined in Eqs.~(\ref{Eq21})-(\ref{Eq24}) as: Fig.~\ref{fig8}(c), LC edge
state; Fig.~\ref{fig8}(d), RM edge state; Fig.~\ref{fig8}(e), RC edge state;
Fig.~\ref{fig8}(f), LM edge state. When the adiabatic approximation holds,
the system will stay in the same eigenstate. Fig.~\ref{fig8}(a) shows how a
LC edge state is adiabatically pumped to the RM edge state during a pumping
cycle, in the mean while Fig.~\ref{fig8}(b) shows how a RC edge state is
adiabatically pumped to the LM edge state within a pumping cycle.

The dynamics of the probability distributions for an open chain with a time
dependent pump sequence defined in Eq.~(\ref{Eq26}) are obtained
numerically. As shown in Fig.~\ref{fig9}(a), the excitation at the first
cavity mode (LC edge state) quickly expands into the bulk, and then the
probability refocuses on the $N$th mechanical mode (RM edge state) at the
end of the first pumping cycle. In the second pumping cycle, the probability
expands into the bulk again and refocuses on the first cavity mode (LC edge
state) at the end of the second pumping cycle. Similarly, as shown in Fig.~%
\ref{fig9}(b), the excitation at the first mechanical mode (LM edge state)
expands into the bulk and refocuses on the $N$th cavity mode (RC edge state)
at the end of the first pumping cycle, and then the probability expands into
the bulk again and refocuses on the first mechanical mode (LM edge state) at
the end of the second pumping cycle. These results are well consistent with
the instantaneous spectrum in the adiabatic limit, as shown in Fig.~\ref%
{fig8}. However, due to the Landau Zener transition occurs at the degenerate
points $\omega t=2n\pi$ ($n$ is positive integer), the periodical behavior of the
probability distribution becomes less well-resolved in the following cycles.

As shown in Figs.~\ref{fig8} and \ref{fig9}, with time dependent pump
sequence defined in Eq.~(\ref{Eq26}), the LC edge state can be pumped
adiabatically to the RM edge state, and the LM edge state can be pumped
adiabatically to the RC edge state, and vice versa. However, the LC edge
state can not be pumped adiabatically to the RC edge state, and the LM edge
state can not be pumped adiabatically to the RM edge state. Now, we consider
another smooth modulation sequence to realizing the adiabatical pumping
between the the LC (LM) edge state and the RC (RM) edge state. The smooth
modulation sequence is defined by
\begin{equation}  \label{Eq27}
\left\{
\begin{array}{c}
u\left( t\right) =\frac{A}{2}\sin \left( \omega t\right) , \\
J\left( t\right) =\mathrm{const.}, \\
v\left( t\right) =A\left[ 1+\cos \left(  \omega t\right) \right] , \\
z\left( t\right) =A\left[ 1-\cos \left( \omega t\right) \right] .%
\end{array}%
\right.
\end{equation}
Here, $u(t)$, $v(t)$ and $z(t)$ are modulated periodically with amplitudes $%
A/2$ and $A$, and frequency $\omega$, while $J(t)$ is maintained constant.

\begin{widetext}
\begin{figure*}[tbp]
\includegraphics[bb=19 456 580 595, width=13 cm, clip]{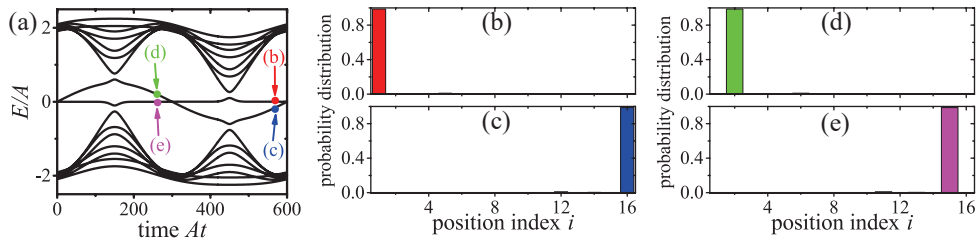}
\caption{(Color online) Instantaneous spectrum of the open chain optomechanical array for $N=8$, $J/A=0.1$, and $\phi=0$.
The time dependent pump sequence is defined in Eq.~(\ref{Eq27}) with $\omega/A=2\pi/600$. (b)-(e) show the probability distributions of the edge states corresponding to the points shown in (a).}
\label{fig10}
\end{figure*}
\end{widetext}

The instantaneous spectrum of the Hamiltonian in Eq.~(\ref{Eq25}) with time
dependent pump sequence defined in Eq.~(\ref{Eq27}) is shown in Fig.~\ref%
{fig10}(a). In Figs.~\ref{fig10}(b)-\ref{fig10}(e), the probability
distributions of eigenstates corresponding to the points shown in Fig.~\ref%
{fig10}(a), are approximately the edge states defined in Eqs.~(\ref{Eq21})-(%
\ref{Eq24}) as: Fig.~\ref{fig10}(b) LC edge state; Fig.~\ref{fig10}(c) RM
edge state; Fig.~\ref{fig10}(d) LM edge state; Fig.~\ref{fig10}(e) RC edge
state. When the adiabatic approximation holds, the system will stay in the
same eigenstate. Fig.~\ref{fig10}(a) shows how a LC edge state is
adiabatically pumped to the RC edge state during a pumping cycle, in the
mean while, a LM edge state is adiabatically pumped to the RM edge state
within a pumping half-cycle.

\begin{figure}[tbp]
\includegraphics[bb=72 393 541 590, width=8.5 cm, clip]{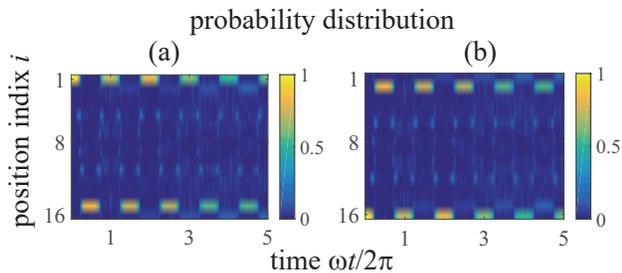}
\caption{(Color online) Time evolution of the probability distribution for
an open chain of $N=8$. The time dependent pump sequence is defined in Eq.~(%
\protect\ref{Eq27}) with $\omega/A=2\pi/600$, (a) $P_{1}(0)=1$; (b) $%
P_{16}(0)=1$. The other parameters are $J/A=0.1$ and $\protect\phi=0$.}
\label{fig11}
\end{figure}

The dynamics of the probability distribution for an open chain with a time
dependent pump sequence defined in Eq.~(\ref{Eq27}) are shown in Fig.~\ref%
{fig11}. In Fig.~\ref{fig11}(a), the excitation at the first cavity mode (LC
edge state) quickly expands into the bulk, and then the probability
refocuses on the $N$th cavity mode (RC edge state) at the end of the first
pumping half-cycle. In the second pumping half-cycle, the probability
expands into the bulk again and refocuses on the first cavity mode (LC edge
state) at the end of the first pumping cycle. Similarly, as shown in Fig.~%
\ref{fig11}(b), the excitation at the first mechanical mode (LM edge state)
expands into the bulk and refocuses on the $N$th mechanical mode (RM edge
state) at the end of the first pumping half-cycle, and then the probability
expands into the bulk again and refocuses on the first mechanical mode (LM
edge state) at the end of the first pumping cycle. These results are
consistent well with the instantaneous spectrum in the adiabatic limit, as
shown in Fig.~\ref{fig10}. However, due to the Landau Zener transition
occurs at the degenerate points $\omega t=n\pi$ ($n$ is positive integer), the
periodical behavior of the probability distribution becomes less
well-resolved in the following cycles.

Overall, with time dependent pump sequence defined in Eq.~(\ref{Eq26}), the
LC edge state can be pumped adiabatically to the RM edge state, and the LM
edge state can be pumped adiabatically to the RC edge state; with time
dependent pump sequence defined in Eq.~(\ref{Eq27}), the LC edge state can
be pumped adiabatically to the RC edge state, and the LM edge state can be
pumped adiabatically to the RM edge state. Therefore, the four edge states
can be pumped from one to another adiabatically with smooth modulation
sequence.

\section{Conclusions}

In summary, we have proposed an implementation of a generalized SSH model
based on optomechanical arrays. This generalized SSH model supports two
distinct nontrivial topological phases, and the transitions between
different phases can be observed by tuning the strengths and phases of the
effective optomechanical interactions.
Dynamic control of the effective optomechanical interactions can be realized by tuning the
strengths and phases of external driving lasers, which allows for dynamic
control of the topological phase transitions. Moreover, four types of edge states can be generated in
the generalized SSH model of an open chain under single-particle excitation,
and the dynamical behaviors of the excitation in the open chain depend on the topological properties under the periodic boundary condition. We show that the edge states can be pumped adiabatically along the optomechanical
arrays by periodically modulating the amplitudes and frequencies of the
driving fields.
Our results can be applied to control the transport of photons and phonons, and the generalized SSH model based on the optomechanical arrays provides us a tunable platform to engineer topological phases for photons
and phonons.

\appendix

\section{Supplemental dispersion relations}

\begin{widetext}
\begin{figure*}[tbp]
\includegraphics[bb=8 331 589 560, width=15 cm, clip]{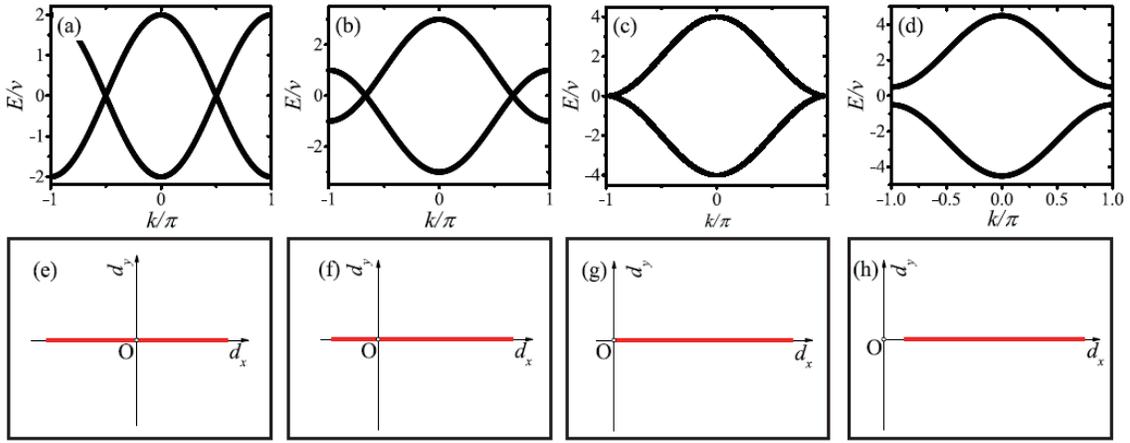}
\caption{(Color online) Dispersion relations of the generalized SSH model with different hopping amplitude $J$: (a) $J=0$; (b) $J/v=1$; (c) $J/v=2$; (d) $J/v=2.5$. (e)-(h) The paths of the endpoint of the vector $\textbf{d}(k)$ corresponding to (a)-(d) are shown on the $d_x,d_y$ plane as the wavenumber is sweeped across the Brillouin zone, $k=0\rightarrow 2\pi$. The other parameters are $z/v=1$ and $\phi=0$.}
\label{figA1}
\end{figure*}

\begin{figure*}[tbp]
\includegraphics[bb=7 322 586 550, width=15 cm, clip]{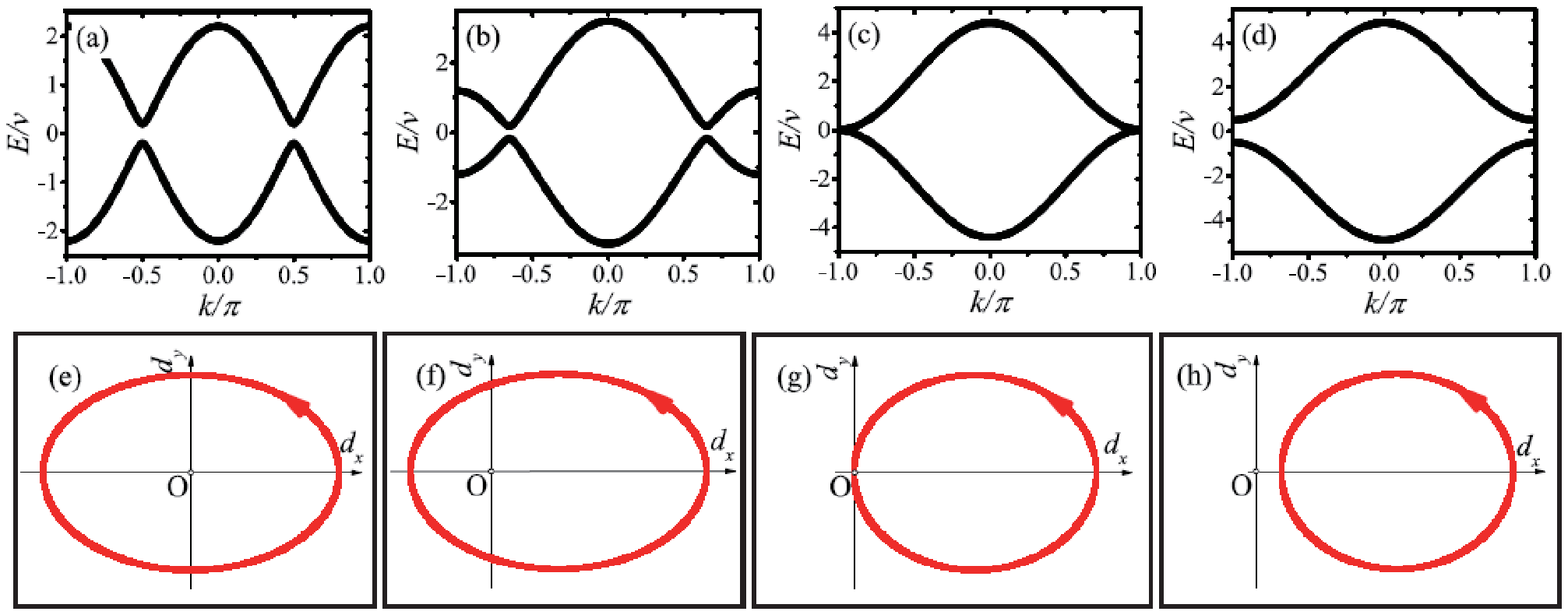}
\caption{(Color online) Dispersion relations of the generalized SSH model with different hopping amplitude $J$: (a) $J=0$; (b) $J/v=1$; (c) $J/v=2.2$; (d) $J/v=2.7$. (e)-(h) The paths of the endpoint of the vector $\textbf{d}(k)$ corresponding to (a)-(d) are shown on the $d_x,d_y$ plane as the wavenumber is sweeped across the Brillouin zone, $k=0\rightarrow 2\pi$. The other parameters are $z/v=1.2$ and $\phi=0$.}
\label{figA2}
\end{figure*}

\begin{figure*}[tbp]
\includegraphics[bb=14 276 578 577, width=11 cm, clip]{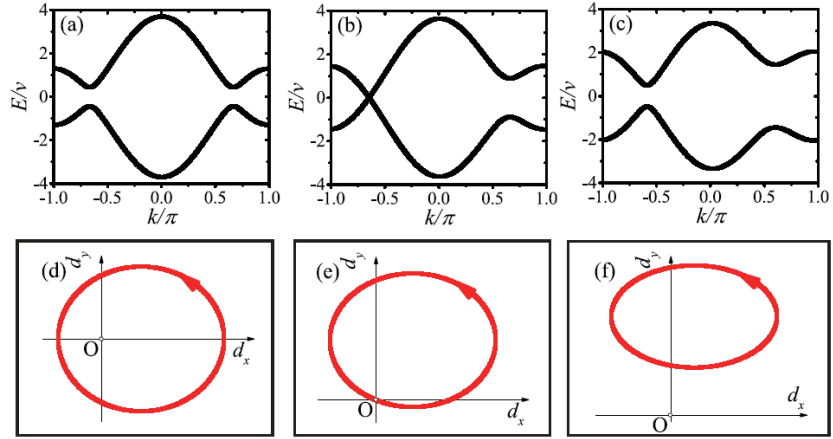}
\caption{(Color online) Dispersion relations of the generalized SSH model for hopping amplitude $J/v=1.2$ with different phase: (a) $\phi=0$; (b) $\phi=\phi_c\approx 0.12$; (c) $\phi=0.3$. (d)-(f) The paths of the endpoint of the vector $\textbf{d}(k)$ corresponding to (a)-(c) are shown on the $d_x,d_y$ plane as the wavenumber is sweeped across the Brillouin zone, $k=0\rightarrow 2\pi$. The other parameter is $z/v=1.5$.}
\label{figA3}
\end{figure*}
\end{widetext}

\vskip 2pc \leftline{\bf Acknowledgement}

X.W.X. is supported by the National Natural Science Foundation of China
(NSFC) under Grants No.11604096 and the Startup Foundation for Doctors of
East China Jiaotong University under Grant No. 26541059. Y.J.Z. is supported
by the China Postdoctoral Science Foundation under grant No. 2017M620945.
A.X.C. is supported by NSFC under Grant No. 11775190. Y.X.L. is supported by
the National Basic Research Program of China(973 Program) under Grant No.
2014CB921401, the Tsinghua University Initiative Scientific Research
Program, and the Tsinghua National Laboratory for Information Science and
Technology (TNList) Cross-discipline Foundation.

\bibliographystyle{apsrev}
\bibliography{ref}

\end{document}